\documentclass[reprint, amsmath,amssymb,aps,]{revtex4-1}

\usepackage{lipsum} 
\usepackage{color} 
\usepackage{graphicx}
\usepackage{dcolumn}
\usepackage{bm}
\usepackage{xcolor}
\usepackage{epstopdf}

\def\bD{\mathbf{D}}
\def\bE{\mathbf{E}}

\def\bH{\mathbf{H}}

\def\bJ{\mathbf{J}}

\def\bP{\mathbf{P}}

\def\bU{\mathbf{U}}

\def\bn{\mathbf{n}}

\def\bp{\mathbf{p}}

\def\bx{\mathbf{x}}

\def\sA{\mathsf{A}}

\def\sD{\mathsf{D}}

\def\sG{\mathsf{G}}

\def\sM{\mathsf{M}}

\def\sR{\mathsf{R}}
\def\sS{\mathsf{S}}
\def\sT{\mathsf{T}}

\def\sV{\mathsf{V}}
\def\sW{\mathsf{W}}

\def\sZ{\mathsf{Z}}

\def\se{\mathsf{e}}
\def\sf{\mathsf{f}}

\def\sq{\mathsf{q}}

\def\sv{\mathsf{v}}

\def\sx{\mathsf{x}}
\def\sy{\mathsf{y}}
\def\sz{\mathsf{z}}

\newcommand{\braket}[2]{\ensuremath{\langle #1 | #2 \rangle}}
\newcommand{\braketWM}[2]{\ensuremath{\langle #1 | #2 \rangle_{\sW \sM}}}

\newcommand{\ket}[1]{\ensuremath{| #1 \rangle}}

\begin{document}

\preprint{APS/123-QED}

\title{Modal Analysis of photonic and plasmonic resonators}
\thanks{This work is part of the research program Good Vibrations with project number 14222, which is (partly) financed by the Dutch Research Counsel (NWO).}%

\author{J\"{o}rn Zimmerling}
 \email{JZimmerl@umich.edu}
\altaffiliation[Also at ]{the Department of Mathematics, University of Michigan, 530 Church Street, Ann Arbor, MI 48109-1043, USA.}
\author{Rob Remis}
 \email{R.F.Remis@TUDelft.NL}
\affiliation{%
 Circuits and Systems Group, Faculty of Electrical Engineering, Mathematics and Computer Science, Delft University of Technology, Mekelweg~4, 2628~CD, Delft, The Netherlands
}%

\date{\today}

\begin{abstract}
\noindent
Quasi-normal modes (QNMs) are ubiquitous throughout photonics and are utilized in a wide variety of applications, but determining these modes remains a formidable task in general. Here we show that by exploiting the structure of Maxwell's equations it is possible to effectively compute QNMs of photonic and plasmonic nanoresonators. The symmetry of Maxwell's equations allows for a reduction to a system of small order via a Lanczos reduction process through which dominant QNMs can be identified. A closed-form reduced-order model for the spontaneous decay (SD) rate of a quantum emitter is also obtained, which does not require an \emph{a priori} QNM expansion of the fields. The model is parametric in wavelength and field expansions in dominant QNMs are determined a posteriori. We demonstrate and validate that QNMs of open resonators and the SD rate of a quantum emitter are accurately predicted. 

\end{abstract}

\pacs{Valid PACS appear here}
\maketitle


\noindent
Optical nanoresonators enable us to confine electromagnetic energy to subwavelength domains and give rise to locally enhanced fields that may stimulate various optical processes in a wide variety of applications and research areas such as biophotonics, optical antennas, and diffraction gratings \cite{Biosensing, OptAntennas, Gratings}. Resonators consisting of metallic nanoparticles that are excited by femtosecond laser pulses are often of particular interest \cite{Femto}, since such resonators allow for the control of light-matter interactions with nanometer and subfemtosecond precision in space and time, respectively, thereby enabling new and exciting applications in cell biology and quantum optics, for example. Moreover, metallic nanoparticles are also often used in resonating structures designed to enhance the SD rate of a quantum emitter that is embedded in such a structure, since this rate depends on the surroundings of the emitter and can be enhanced by an electromagnetic resonance (Purcell effect). The spontaneous decay of a quantum emitter is a purely quantum mechanical effect, but can be computed classically in the so-called weak-coupling regime \cite{Novotny&Hecht}. Specifically, with $\gamma$ denoting the decay rate of the emitter in the resonator configuration of interest and $\gamma_0$ the decay rate of the same emitter in a reference medium, $\gamma/\gamma_0 = P/P_0$, where $P$ and $P_0$ are the time-averaged powers radiated by an electric dipole positioned at the location of the emitter in the resonator configuration and reference medium, respectively. Explicitly, for an emitter located at $\bx=\bx_{\text{S}}$ and an electric dipole of the form $\bJ^{\text{ext}}=\partial_t \bp(t) \delta(\bx-\bx_{\text{S}})$ with dipole moment $\bp(t)=p(t)\bn_{\text{s}}$, $p(t)=|\bp(t)|$, and $\bn_{\text{s}}$ a unit vector, we have in steady-state $\hat{\bJ}^{\text{ext}}=-\text{i}\omega \hat{p}(\omega) \delta(\bx-\bx_{\text{S}}) \bn_{\text{s}}$ and the time-averaged radiated power is given by
\begin{align}
\label{eq:en_disp}
P(\omega)
=
\frac{\omega}{2}
\text{Im}
\left[
\hat{p}^{\ast}(\omega)\, \hat{\bE}(\bx_{\text{S}},\omega) \cdot \bn_{\text{s}}
\right].
\end{align}
To evaluate this power over a frequency or wavelength interval of interest, the electric field strength at the dipole location is required for all frequencies belonging to this interval.

To investigate what local field or decay rate enhancements can be realized, a modal analysis of a resonating structure is typically carried out. For open resonator structures these modes are called Quasi Normal Modes or QNMs and are characteristic of the structure at hand and independent of the excitation. An external source (or incident field) determines what resonant modes are actually excited, while the contribution of these excited modes to a measured field response is determined by the receiver. In open resonant structures, typically only a small number of QNMs are necessary to accurately model measured field responses \cite{LalanneOverviewArt,PhysRevLettLalanne,LineINtegralMethod2018,Gras19,Lalanne_19,Zolla:18} and in SD rate computations the source and receiver location actually coincide, since the electric field strength at the source (dipole) location is required to determine the radiated power (see Eq.~(\ref{eq:en_disp})).

In this letter we show that by exploiting the symmetry of the first-order Maxwell system, it is possible to efficiently determine QNMs of open resonating structures consisting of dispersive metallic nanoparticles. In addition, we show that the SD rate can be computed without any \emph{a priori} mode selection, that is, the decay rate can be computed without an explicit mode expansion of the fields as is more commonly done in decay rate computations (see, e.g.~\cite{PhysRevLettLalanne}).

To describe the reaction of a metallic nanoparticle to the presence of an electromagnetic field, we write the electric displacement vector in Maxwell's equations as $\hat{\bD}=\varepsilon\hat{\bE} + \hat{\bP}=\varepsilon_{\text{c}}(\omega) \hat{\bE}$ with $\varepsilon=\varepsilon_0 \varepsilon_{\infty}$, where $\varepsilon_{\infty}$ is the instantaneous (high-frequency) permittivity and a polarization vector $\hat{\bP}$ that is related to the electric field strength via the generic constitutive relation $-\omega^2 \hat{\bP} -\text{i}\omega \beta_2 \hat{\bP} + \beta_1 \hat{\bP} = \beta_0 \hat{\bE}$, where the coefficients $\beta_i$ determine what type of relaxation is considered (Drude, Lorentz). For a Drude model, for example, we have $\beta_0=\varepsilon_0 \omega_{\text{p}}^2$, $\beta_1=0$, and $\beta_2=\gamma_{\text{p}}$, where $\omega_{\text{p}}$ is the volume plasma frequency and $\gamma_{\text{p}}$ the collision frequency of the metal.

Introducing the auxiliary field variable $\hat{\bU} = \text{i}\omega \hat{\bP}$, we can write the above constitutive relation and Maxwell's equations in the consistent first-order form \cite{Zimmerling2016_2}
\begin{equation}
\label{eq:maxwell}
\begin{bmatrix}
-i \omega \varepsilon &&-1&-\nabla \times\\
&-i \omega  &1&&\\
\beta_0&-\beta_1&\beta_2-i\omega&\\
\nabla \times&&&- i \omega \mu
\end{bmatrix}
\begin{bmatrix}
\hat\bE\\
\hat\bP\\
\hat\bU\\
\hat\bH
\end{bmatrix}
=
-
\begin{bmatrix}
	\hat\bJ^{\text{ext}} \\
	\mathbf{0} \\
	\mathbf{0} \\
	\mathbf{0}
\end{bmatrix},
\end{equation}
which can be written as $\left(\mathcal{D} + \mathcal{S} -\text{i}\omega \mathcal{M}\right)\hat{\mathcal{F}} = -\hat{\mathcal{Q}}$, where $\mathcal{S}$ and $\mathcal{M}$ are medium matrices, and the curl operators of Maxwell's equations are contained in the spatial differentiation operator $\mathcal{D}$. The electromagnetic field quantities and external sources are collected in the field vector~$\hat{\mathcal{F}}$ and source vector $\hat{\mathcal{Q}}$, respectively. For most external sources used in practice (electric dipole, for example), the frequency dependence of the source can be factored out and we write $\hat{\mathcal{Q}}=\hat{p}(\omega) \mathcal{Q}'$, where $\hat{p}(\omega)$ is the source wavelet and $\mathcal{Q}'$ is frequency independent.

We note that the partial differential operator in Eq.~(\ref{eq:maxwell}) can be symmetrized by scaling the second row with $\beta_1\beta_0^{-1}$, the third row with $-\beta_0^{-1}$, and the fourth row by $-1$. The efficiency of our method is based upon this symmetry. Furthermore, measured (causal) material behavior can be modeled using this formulation by fitting a rational function representation (i.e. a multipole expansion consisting of a superposition of Lorentz and Drude models) for the complex permittivity to permittivity measurements. This leads to the introduction of multiple auxiliary field variables and the resulting system can be symmetrized in a similar manner as described above.

To carry out a modal analysis of arbitrarily-shaped open resonators, we discretize the first-order Maxwell system in space using a staggered finite-difference Yee mesh. We discretize on such a mesh, since it can be shown that the discretization procedure is mimetic, that is, it is structure preserving and conservation laws and important physical symmetry properties of Maxwell's equations (symmetry related to energy conservation or symmetry related to reciprocity, for example) have a counterpart after discretization  \cite{ChewFD, Zimmerling2016_2}. Other discretization schemes (finite elements, for example) can also be used, of course, so long as these schemes are mimetic as well.

In addition, radiation towards infinity has to be taken into account, since we are interested in open nanoresonators. Typically, this is realized by surrounding the domain of interest by a so-called Perfectly Matched Layer (PML) \cite{Berenger} in which the spatial coordinates are stretched using frequency \emph{dependent} stretching functions \cite{Chew1994}. However, a disadvantage of such an approach is that in two- and three-dimensional problems this leads to nonlinear eigenvalue problems that need to be solved to find dominant QNMs. Therefore, our approach is to apply the PML technique of \cite{2013Remis,PML2016}, which uses complex spatial step sizes to realize a perfectly matched layer, which do \emph{not} explicitly depend on frequency and leads to linear eigenproblems. Incorporating this PML technique into our spatial discretization scheme then leads to the discretized first-order Maxwell system
\begin{equation}
\label{eq:maxwell_d}
\left(
\mathsf{D} + \mathsf{S} -\text{i}\omega \mathsf{M}
\right)
\hat{\mathsf{f}}_{\text{cs}} = -\hat{p}(\omega)\sq',
\end{equation}
where $\sD$ contains the discretized curl operators, $\sS$ and $\sM$ are the discretized medium matrices, and $\hat{\sf}_{\text{cs}}$ and $\sq'$ are the discretized field and source vector, respectively. The above system is not conjugate-symmetric with respect to frequency and its time-domain counterpart is unstable due to the application of a frequency-independent PML. However,  conjugate-symmetric frequency-domain field approximations can be obtained from the above system as \cite{2013Remis}
\begin{equation}
\label{eq:field_approx}
\hat{\mathsf{f}}(\omega) = -\hat{p}(\omega)\hat{\mathsf{G}}(\mathsf{A},\omega) \sq,
\end{equation}
where $\mathsf{A}=\mathsf{M}^{-1}(\mathsf{D}+ \mathsf{S})$ is the first-order Maxwell system matrix, $\sq= \sM^{-1} \sq'$ is the scaled source vector, and $\hat{\sG}(\sA,\omega) = \hat{\mathsf{R}}(\mathsf{A},\omega) + \hat{\mathsf{R}}^{\ast}(\mathsf{A},-\omega)$ is the Green's tensor of the configuration with $\hat{\mathsf{R}}$ the filtered resolvent of matrix~$\mathsf{A}$ given by $\hat{\mathsf{R}}(\mathsf{A},\omega) =\chi(\mathsf{A})(\mathsf{A} - \text{i}\omega \mathsf{I})^{-1},$ in which $\chi(z)$ is the complex Heaviside unit step function defined as $\chi(z)=1$ for $\text{Re}(z)>0$ and $\chi(z)=0$ for $\text{Re}(z)<0$. Note that $\hat{\mathsf{f}}(\omega)$ is conjugate-symmetric, that is, it satisfies $\hat{\mathsf{f}}^{\ast}(\omega)=\hat{\mathsf{f}}(-\omega)$, provided that $\hat{p}$ is conjugate-symmetric.

For practical three-dimensional problems direct evaluation of Eq.~(\ref{eq:field_approx}) is usually not feasible, since the order $n$ of the Maxwell system matrix~$\sA$ is simply too large (in 3D, typically $n=O(10^{6 - 7})$). It can be shown, however, that matrix~$\sA$ satisfies a particular symmetry property that allows for efficient Lanczos model-order reduction.  In particular, defining the bilinear form $\braketWM{\sx}{\sy} = \sx^{T} \sM \sW \sy = \sx^{T} \sW \sM \sy,$
for vectors from $\mathbb{C}^{n}$, where $\sW$ is a specific step size matrix \cite{Zimmerling2016_2,Zimmerling2016_1}, it can be shown that $\braketWM{\sA \sx}{ \sy }= \braketWM{\sx}{ \sA\sy }$ for all vectors $\sx,\sy \in \mathbb{C}^{n}$. Moreover, the bilinear form $\braketWM{\sf}{\sf}$ is a discrete approximation of the integral
\begin{align}
\label{eq:lagrangian}
\begin{split}
\mathcal{L} &= \int \varepsilon \hat{\bE}^2   + \beta_1\beta_0^{-1} \hat{\bP}^2 - \beta_0^{-1}  \hat{\bU}^2 - \mu \hat{\bH}^2 \text{d}V \\
&= \int \hat{\bE} \cdot
\frac{\partial \omega \varepsilon_{\text{c}}(\omega)}{\partial \omega} \cdot \hat{\bE} - \mu \hat{\bH} \cdot \hat{\bH} \, \text{d}V
\end{split}
\end{align}
which in the literature  is used to normalize QNMs \cite{PhysRevLettLalanne}.

In 1931, Krylov~\cite{Krylov1931} used what are now called polynomial Krylov subspaces in his analysis of oscillations of mechanical systems (e.g. ships). Here, the symmetry of matrix~$\sA$ allows us to follow a similar approach. Specifically, the symmetry of $\sA$ can be used to reduce this matrix to tridiagonal form using a three-term Lanczos-type recurrence relation \cite{Zimmerling2016_2,Zimmerling2016_1}. Carrying out $m$ steps of this reduction process, we obtain the  decomposition
\begin{equation}
\label{eq:Lanczos_decomp}
\sA \sV_{m} = \sV_{m} \sT_{m} + \beta_{m+1} \sv_{m+1} \se_{m}^{T},
\end{equation}
where $\sT_m$ is a tridiagonal matrix of order $m \ll n$ containing the Lanczos recurrence coefficients and $\sV_m$ is a tall $n$-by-$m$ matrix with a column partitioning $\sV_m=(\sv_1, \sv_2,...,\sv_m)$. The columns of matrix $\sV_m$ are referred to as Lanczos vectors, which are taken to be quasi-orthonormal i.e., $\braketWM{ \sv_i}{\sv_j } = \delta_{ij}$, where $\delta_{ij}$ is the Kronecker delta. Furthermore, $\beta_{m+1}$ in Eq.~(\ref{eq:Lanczos_decomp}) is a Lanczos recurrence coefficient and $\se_m$ is the $m$th canonical basis vector. To find an approximate spectrum of the Maxwell system matrix~$\sA$, the Lanczos reduction process can be started with any (randomly generated) starting vector $\sv_1$ satisfying $\braketWM{\sv_1}{\sv_1} =1$. If, however, modes excited by a given external source are of interest (as in SD rate computations, for example) then we take $\sv_1= \braketWM{\sq}{\sq}^{-1/2} \sq$ as a starting vector in the reduction process.

The Lanczos decomposition of Eq.~(\ref{eq:Lanczos_decomp}) serves as a starting point for our modal analysis and SD rate computations. First, as is well known~\cite{Golub&VanLoan}, the decomposition can be used to find approximate QNMs of the open resonator system. Specifically, if $(\theta_j^{[m]},\sz_j^{[m]})$ is an eigenpair of the reduced matrix~$\sT_m$ then postmultiplication of (\ref{eq:Lanczos_decomp}) by $\sz_j^{[m]}$ shows that $(\theta_j^{[m]},\sV_m\sz_j^{[m]})$ is an approximate eigenpair of $\sA$ with a residual given by $\beta_{m+1}\langle \se_{m}|\sz^{[m]}\rangle |\sv_{m+1}\rangle$. Converged QNMs $\sy_j=\sV_m\sz_j^{[m]}$ can be identified by computing the norm of this residual. Note that normalizing the eigenvectors $\sz_j^{[m]}$ of $\sT_m$ such that $\braket{\sz_j^{[m]}}{\sz_i^{[m]}}=\delta_{ij}$ ensures that the approximate QNMs $\sy_j$ are normalized with respect to the bilinear form~(\ref{eq:lagrangian}), i.e. $\braketWM{\sy_j}{\sy_i}=\delta_{ij}$.

Second, for a given external source~$\sq$ the decomposition can be used to construct the reduced-order model (ROM) \cite{Zimmerling2016_2,2013Remis}
\begin{align}
\label{eq:ROM}
\begin{split}
\hat{\mathsf{f}}_m(\omega) &= 
\text{i} \omega \hat{p}(\omega) \braketWM{\sq}{\sq}^{1/2} \times \\
&\left[
\sV_m \hat{\sR}(\sT_m,\omega) \ket{\se_1} +  
\sV_m^{\ast} \hat{\sR}^\ast(\sT_m,-\omega) \ket{\se_1}
\right],
\end{split}
\end{align}
which gives an approximation of the three-dimensional field of order $m$. In SD rate computations, however, only the projection of the electric field onto the direction of the dipole moment at the dipole location is of interest. For this projection, we have $\hat{\bE}(\bx_{\text{S}},\omega) \cdot \bn_{\text{s}} \approx \braketWM{\hat{\sf}_m(\omega)}{\sq}$ and substitution in Eq.~(\ref{eq:en_disp}) gives the ROM for the radiated power 
\begin{equation}
\label{eq:SD_rate_ROM}
P_m(\omega)
= P_\text{a}\,
\text{Re}
\left[
\braket{\se_1 }{ \hat{\sG}(\sT_m,\omega)| \se_1}
\right]
\end{equation}
with $P_\text{a}=0.5\, \omega^2 |\hat{p}|^2 \braketWM{\sq}{\sq}$. Only filtered resolvents of the reduced tridiagonal matrix $\sT_m$ need to be computed to evaluate this power over a complete frequency (wavelength) interval of interest and no \emph{a priori} expansion of the fields in QNMs in required. Explicitly, assuming that $\sT_m$ can be diagonalized and arranging its eigenvectors as columns in matrix $\sZ_m=(\sz_1^{[m]},\sz_2^{[m]},...,\sz_m^{[m]})$, we have 
\begin{equation}
\label{eq:SD_rate_ROM2}
P_m(\omega)
= P_\text{a}
\text{Re}
\left[
\sum_{k=1}^m
w_k^2 \hat{\sR}(\theta_k^{[m]},\omega) + (w_k^\ast)^2 \hat{\sR}^\ast(\theta_k^{[m]},-\omega)
\right], 
\end{equation}
where $w_k$ is the $k$th element of $|\sZ_m^T \se_1 \rangle$. As mentioned above, converged QNMs can be identified by computing the residual of the approximate QNMs and their contribution to the radiated power $P_m(\omega)$ can be determined using the spectral expansion of Eq.~(\ref{eq:SD_rate_ROM2}). With $\mathcal{I}^{\text{QNM}}$ denoting the index set of converged QNMs that contribute to the radiated power we then arrive at a low order QNM expansion by replacing the sum in Eq.~(\ref{eq:SD_rate_ROM2}) by a sum over all $k \in \mathcal{I}^{\text{QNM}}$. In other words, the Lanczos decomposition allows us to determine \emph{a posteriori} which converged QNMs actually contribute to the radiated power and ultimately the SD rate of the quantum emitter.

\begin{figure}[t]
\begin{picture}(242,188)
\put(0,0){\includegraphics[trim=15mm 11mm 10mm 5mm,clip,width=\columnwidth]{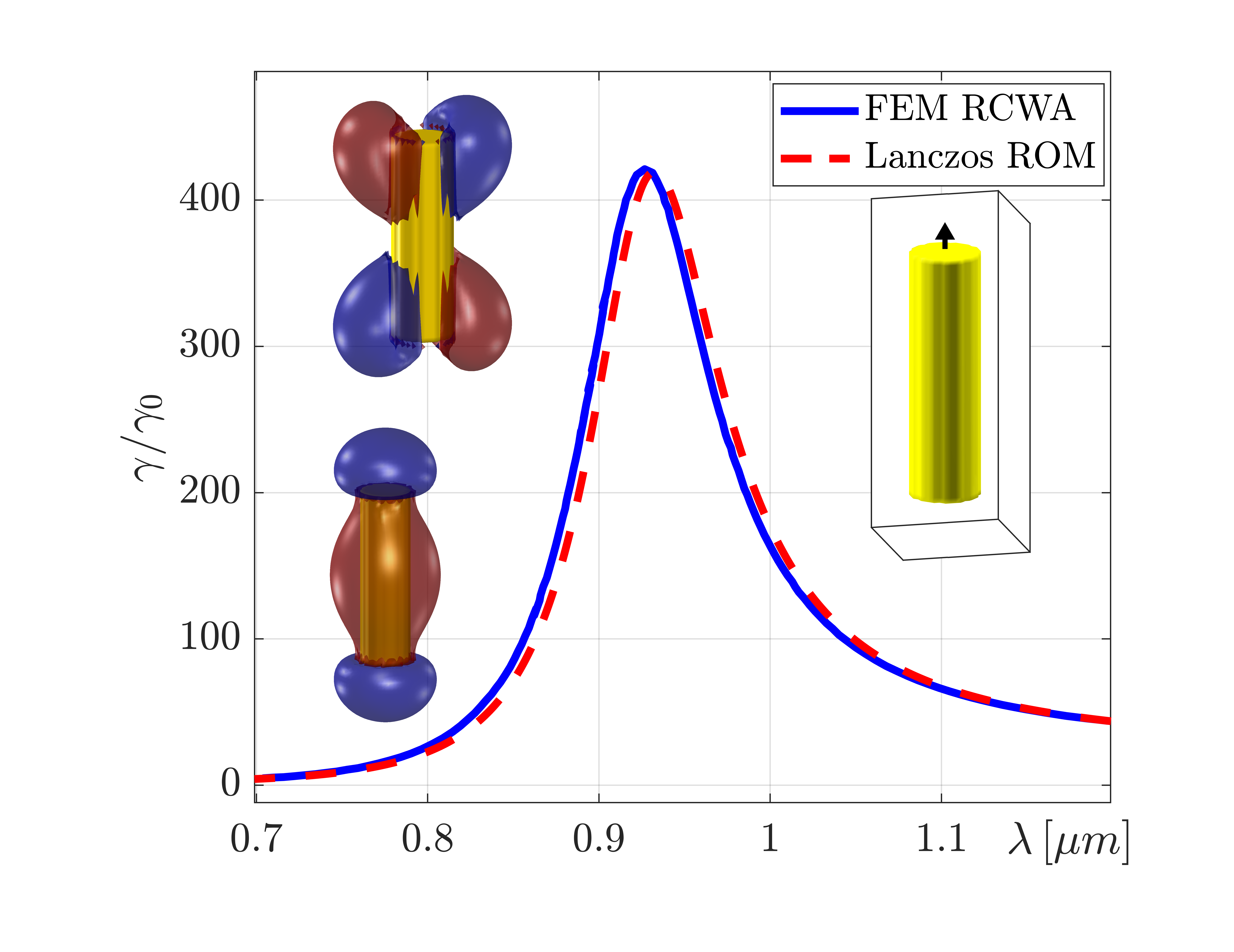}}
\put(220,145){\normalsize (a)}
\put(38,175){\normalsize (b)}
\put(38,95){\normalsize (c)}
\put(171,119){\small x}
\put(177,69.5){\small y}
\put(205,68){\small z}
\end{picture}
\caption{\label{fig:Validation} { Purcell factor of a quantum emitter (arrow) centered 10~nm above a $30~\text{nm} \times100~\text{nm}$ nanorod computed using the FEM-RCWA method \cite{PhysRevLettLalanne} (solid line) and the Lanczos ROM (dashed line). (a)~Simulated configuration, (b)~isosurface plots of $\text{Re}(\hat{E}_z)$~(b) and $\text{Re}(\hat{E}_x)$~(c) of the dominant QNM with a wavelength $\lambda=926 + 47\text{i}$~nm.} }
\end{figure}
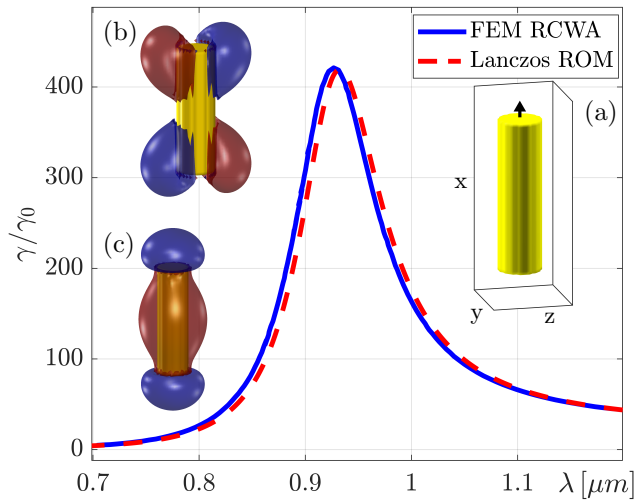

\begin{figure*}[thp]
\begin{picture}(500,105)
\put(0,0){\includegraphics[trim=0mm 0mm 20mm 5mm,clip,width=\textwidth]{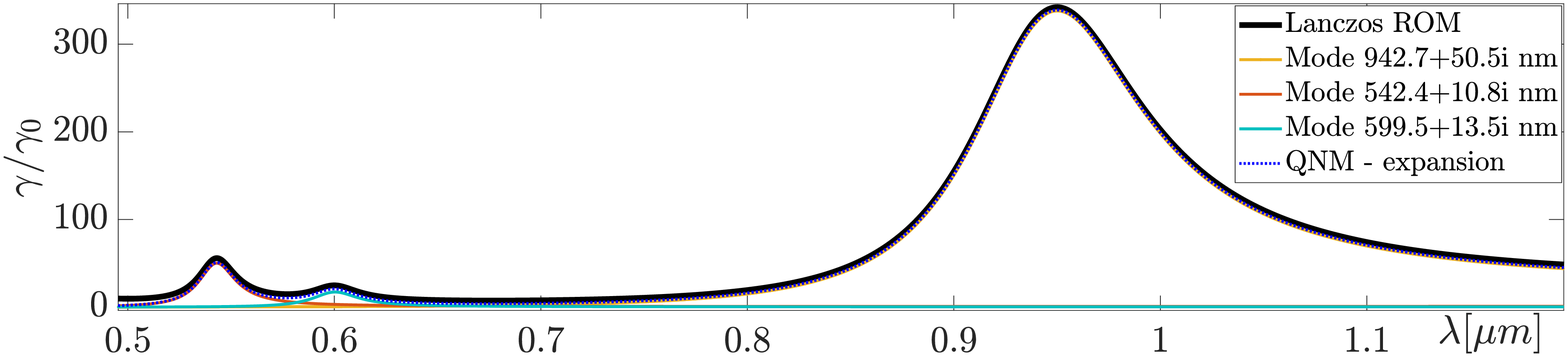}}
\put(142,20){\includegraphics[trim=0mm 0mm 0mm 0mm,clip,width=0.235\textwidth]{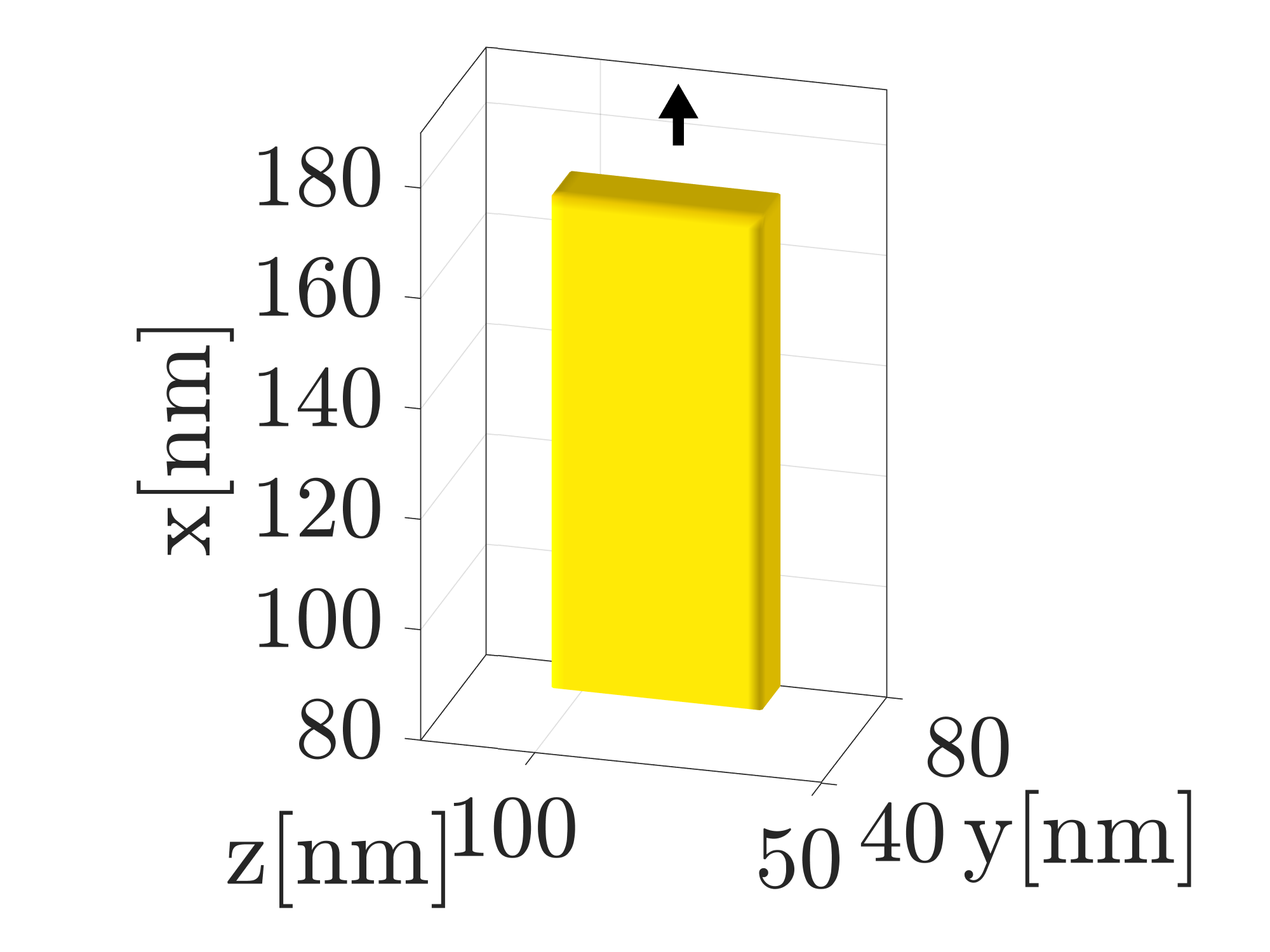}}
\put(39,-10){\includegraphics[trim=80mm 40mm 80mm 35mm,clip,width=0.32\textwidth]{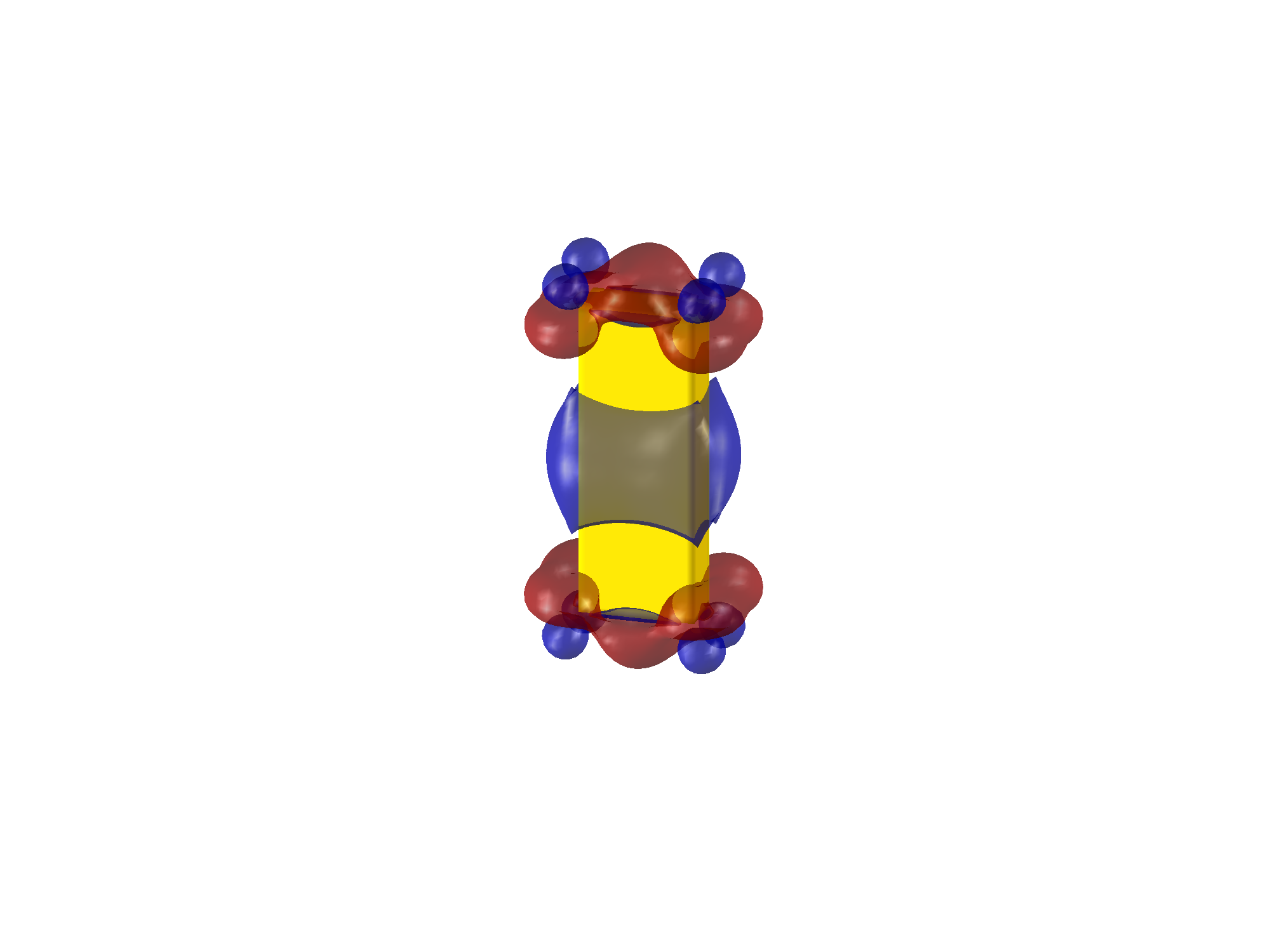}}
\put(-10,0){\includegraphics[trim=80mm 40mm 80mm 35mm,clip,width=0.32\textwidth]{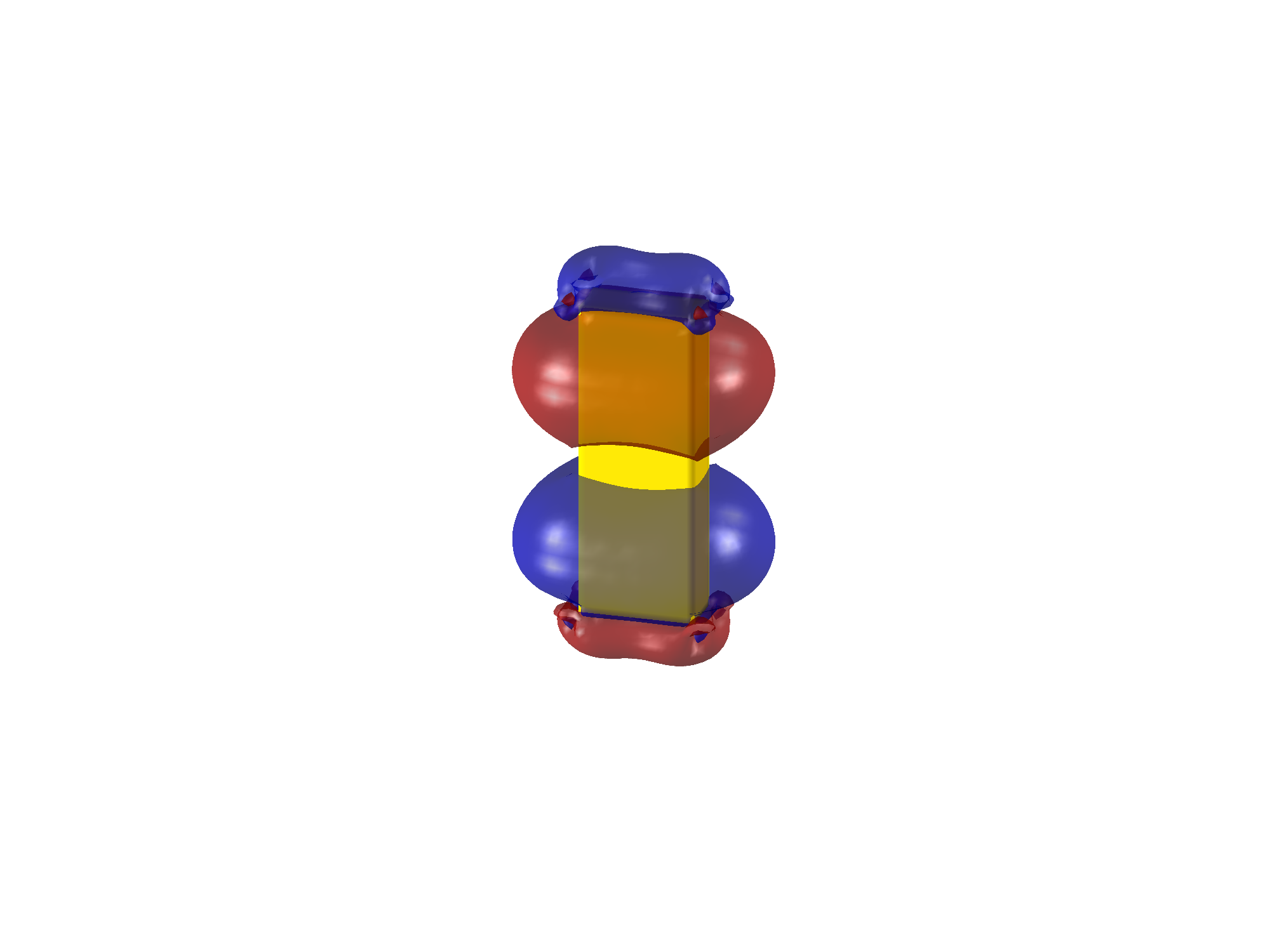}}
\put(255,-20){\includegraphics[trim=80mm 40mm 80mm 35mm,clip,width=0.32\textwidth]{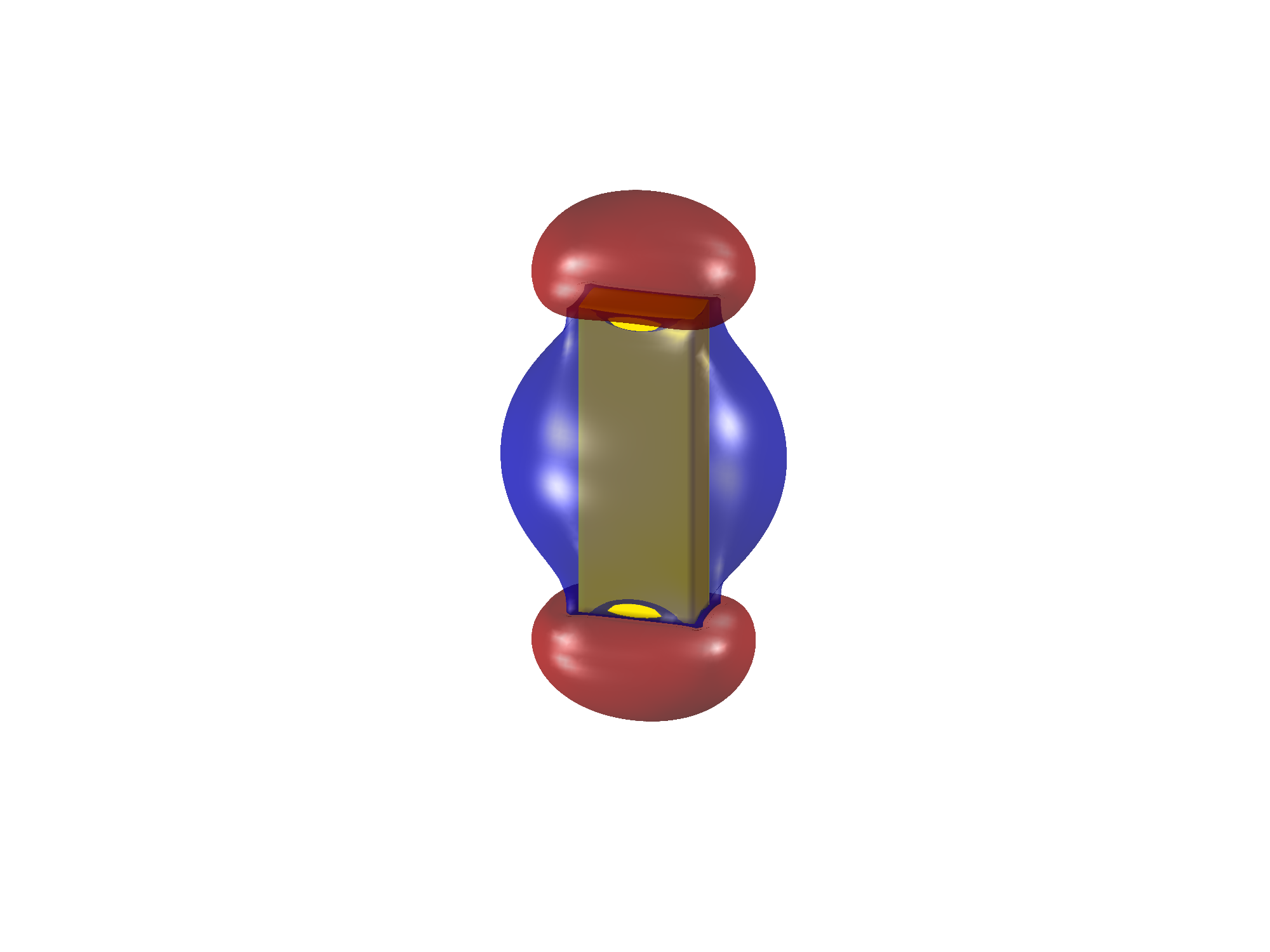}}
\put(42,100){\normalsize (a)}
\put(125,100){\normalsize (b)}
\put(225,100){\normalsize (c)}
\put(300,23){\normalsize (d)}

\end{picture}
\caption{\label{fig:singlePlate}  Purcell factor of a quantum emitter (arrow) located 10~nm above a $102~\text{nm} \times 40~\text{nm} \times 20~\text{nm}$ nanoplate. The Purcell factor is computed using Lanczos reduction (Eq.~(\ref{eq:SD_rate_ROM})) and an expansion in the three most dominant QNMs. The real part of the $\hat{E}_x$ field of the three dominant QNMs is depicted along with their individual contribution to the SD rate. (a) $\text{Re}(\hat{E}_x)$ of the QNM with $\lambda=542.4+10.8\text{i}$~nm. (b) $\text{Re}(\hat{E}_x)$ of the QNM with $\lambda=599.5+13.5\text{i}$~nm. (c) Simulated configuration. (d) $\text{Re}(\hat{E}_x)$ of the QNM with $\lambda=942.7+50.5\text{i}$~nm.}
\end{figure*}

%
%
To validate the presented approach, we compute the Purcell factor of a golden nanorod that has been considered in the literature before~\cite{PhysRevLettLalanne}. The configuration consists of a vertically oriented dipole centered 10~nm above a $30~\text{nm}\times100~\text{nm}$ golden nanorod, embedded in a dispersionless background material with relative permittivity $\varepsilon_\text{r}=2.25$. A Drude model is used as a dispersion model for gold with a plasma frequency  $\omega_\text{p}=1.26 \cdot 10^{16}$~Hz and a collision frequency $\gamma_{\text{p}}=1.41\cdot 10^{14}$~Hz. This dispersion model is used throughout this letter. Finally, we mention that since our reduction framework is designed for arbitrarily-shaped nanoresonators, we do not make use of any rotational symmetry. 

In Fig.~\ref{fig:Validation} the computed Purcell factor over a complete wavelength interval of interest is shown. The solid line signifies the result obtained with the FEM-RCWA method~\cite{PhysRevLettLalanne}, while the dashed line shows the converged reduced-order model response obtained via Lanczos reduction. The computed enhancement factors of both methods are in good agreement with each other. The unreduced Maxwell system has an order of $n=8.6$ million, while the order of the converged reduced system is $m=4500$. Dominant QNMs can be identified from the spectrum of the reduction matrix~$\sT_{4500}$. For this configuration, it turns out that essentially only a single QNM with a complex resonance wavelength of $\lambda= 926 + 47\text{i}$~nm contributes to the SD rate over the considered wavelength interval. Higher order QNMs only contribute to the Purcell factor for wavelengths smaller than 600~nm. Finally, isosurface plots of $\text{Re}(\hat{E}_z)$ and $\text{Re}(\hat{E}_x)$ of the dominant QNM as computed via Lanczos reduction are shown in Figs.~\ref{fig:Validation}(b) and \ref{fig:Validation}(c), respectively, where the red and blue surfaces have opposite signs. The isosurface has been chosen to best visualize the field distribution.

%
%
\begin{figure*}[bht]
\begin{picture}(500,115)
\put(0,0){\includegraphics[trim=35mm 0mm 45mm 0mm,clip,width=0.195\textwidth]{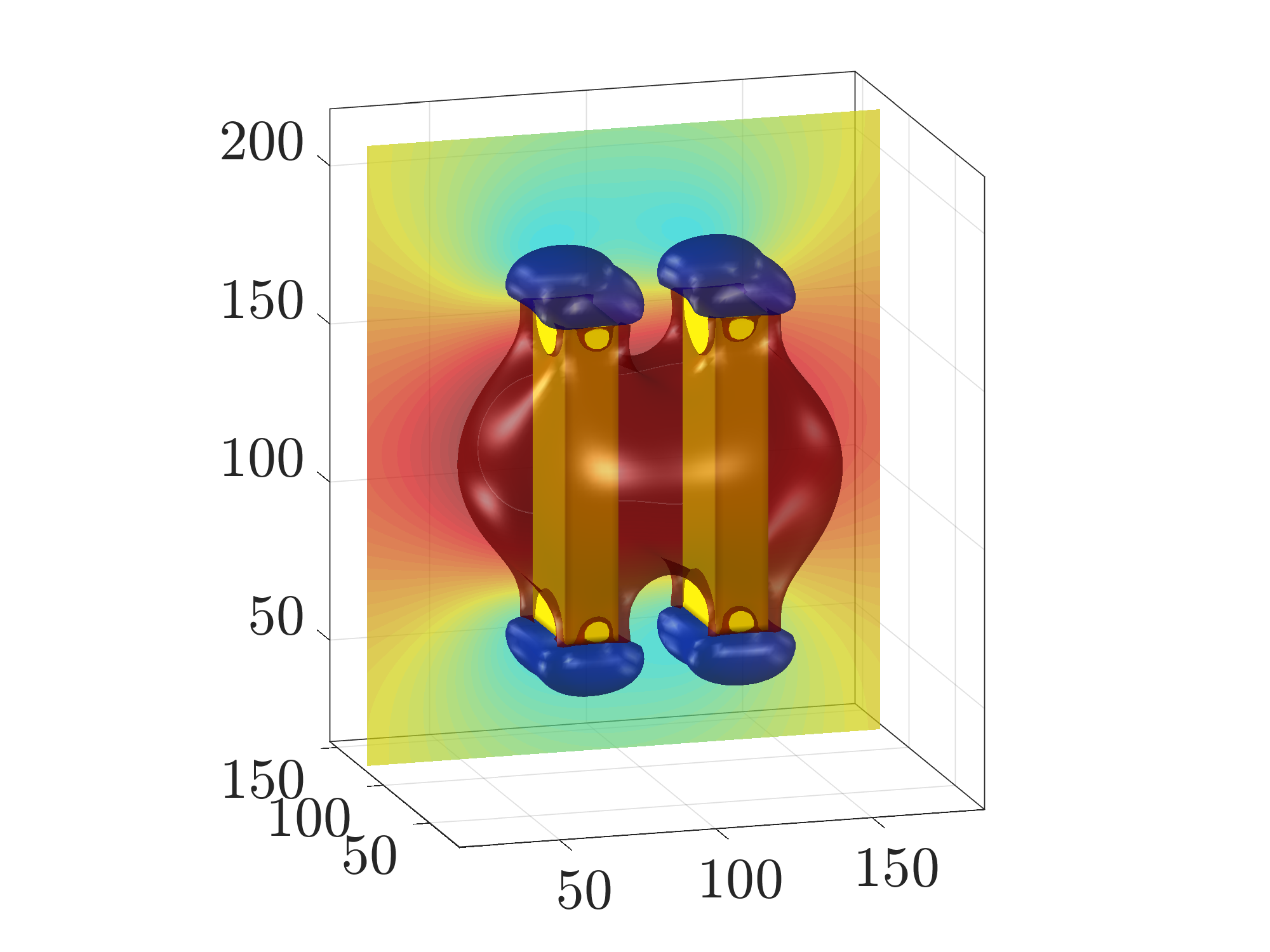}}
\put(101,0){\includegraphics[trim=35mm 0mm 45mm 0mm,clip,width=0.195\textwidth]{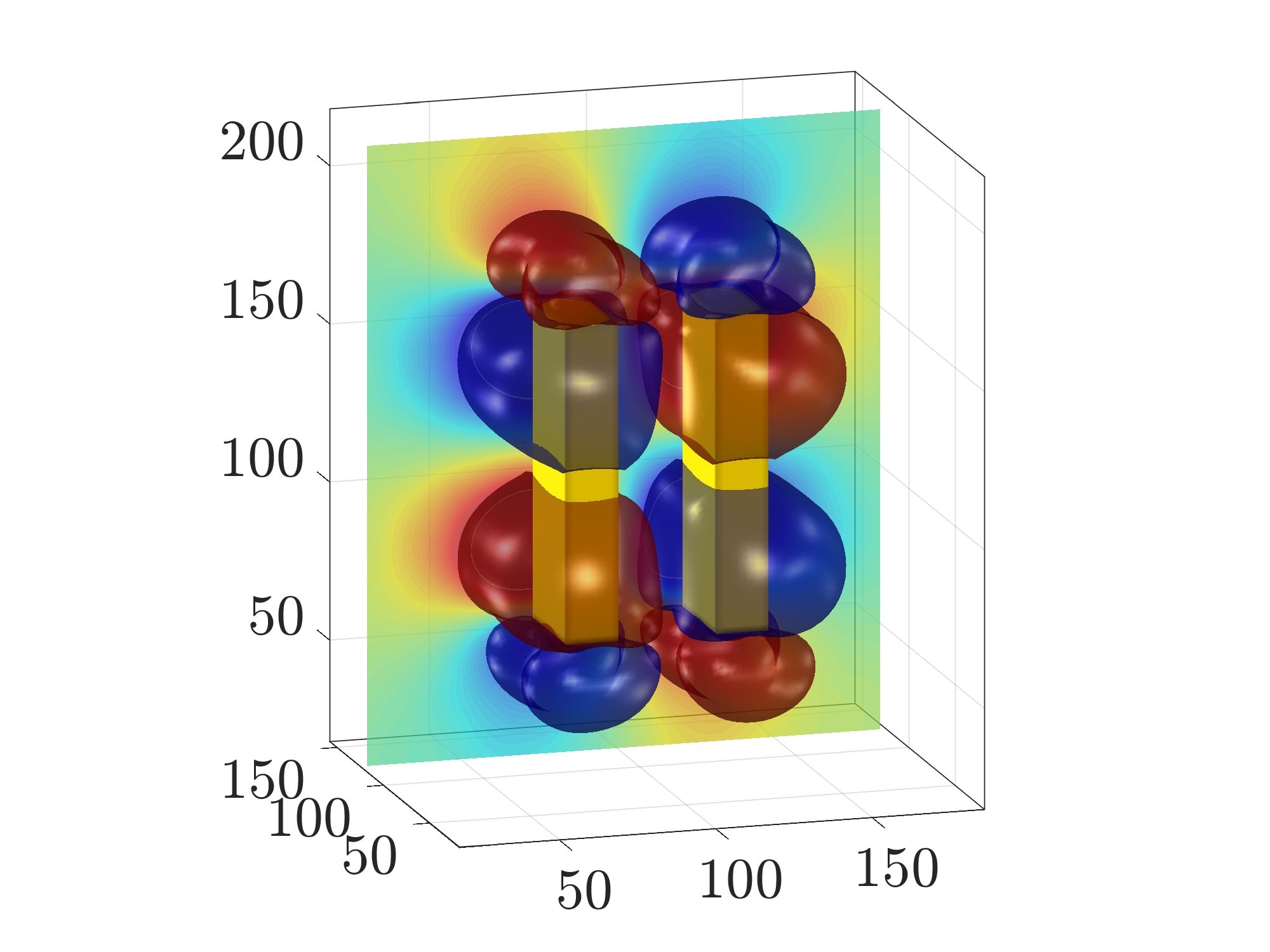}}
\put(202,0){\includegraphics[trim=35mm 0mm 45mm 0mm,clip,width=0.195\textwidth]{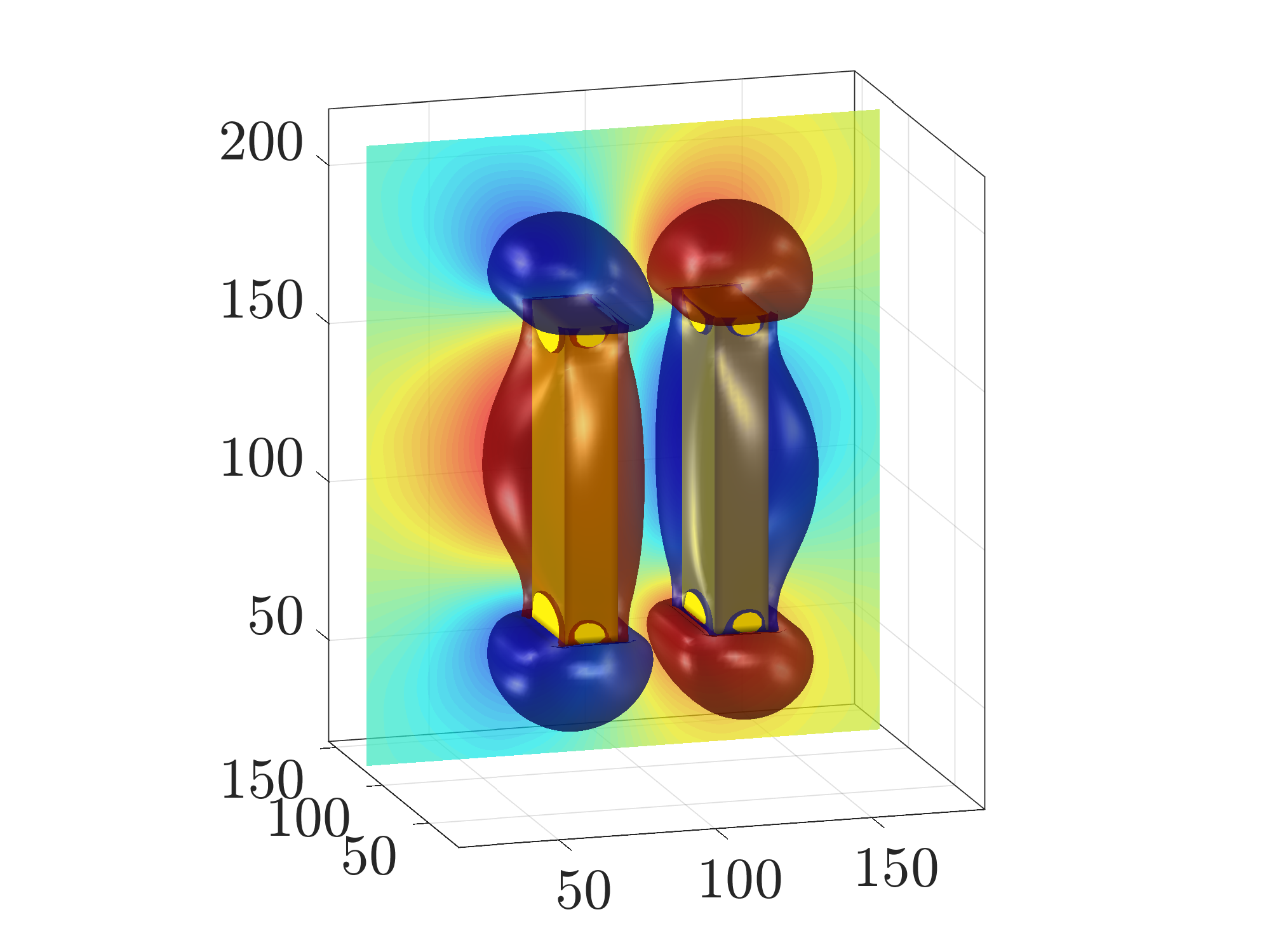}}
\put(303,0){\includegraphics[trim=35mm 0mm 45mm 0mm,clip,width=0.195\textwidth]{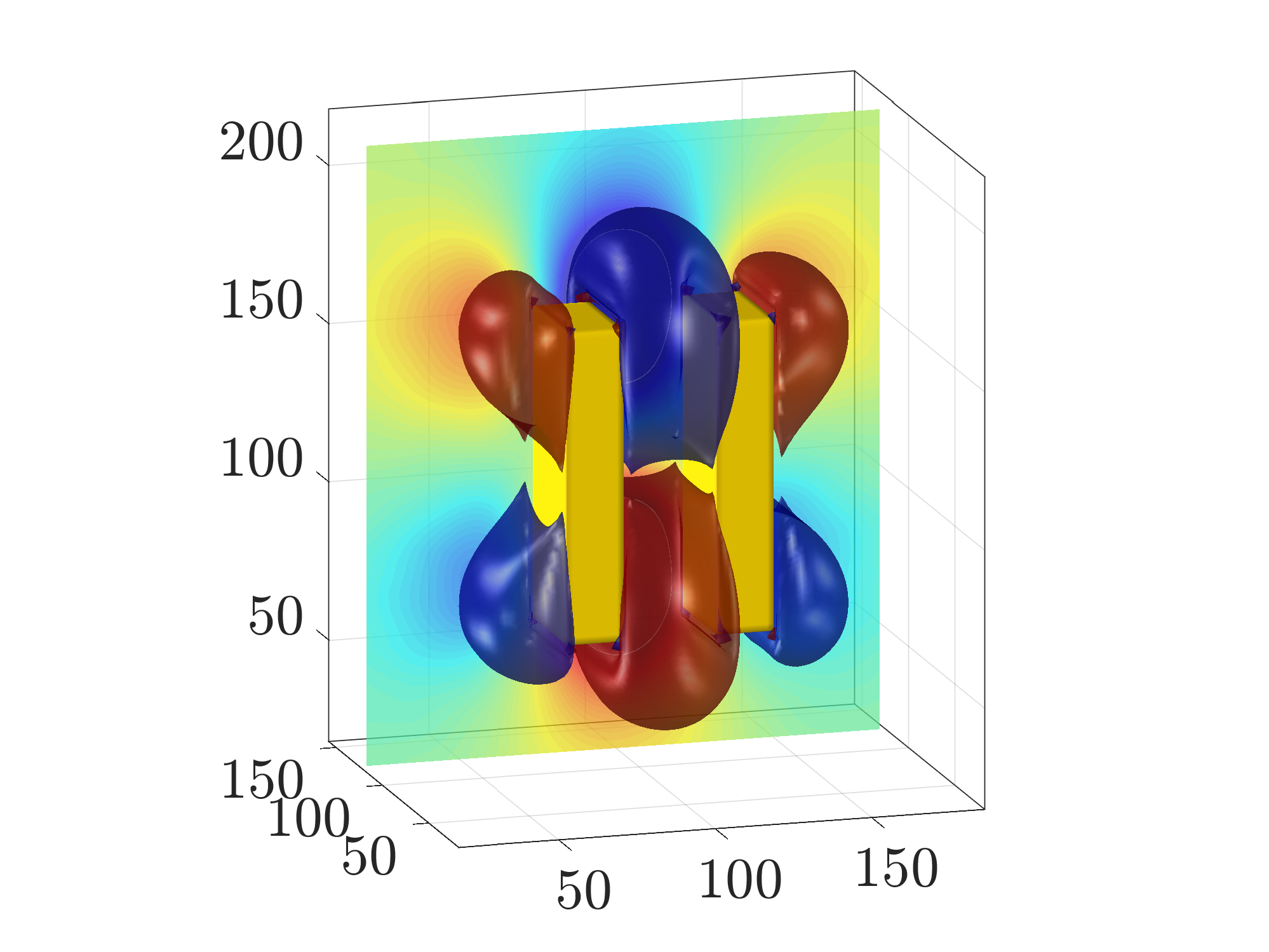}}
\put(404,0){\includegraphics[trim=35mm 0mm 45mm 0mm,clip,width=0.195\textwidth]{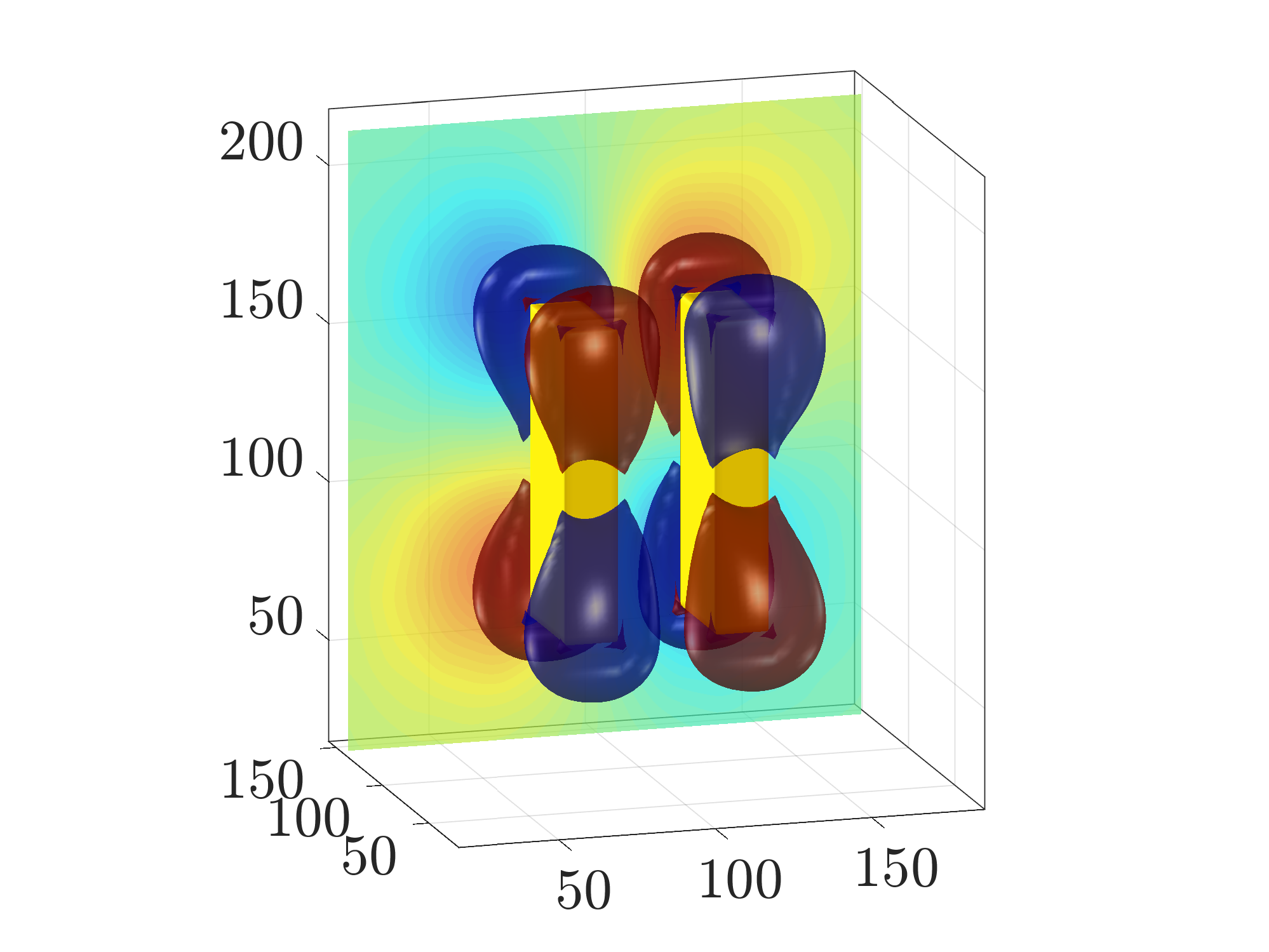}}
\put( 5,113){\large (a)}
\put(106,113){\large (b)}
\put(207,113){\large (c)}
\put(308,113){\large (d)}
\put(409,113){\large (e)}
\put(5,8){\scriptsize z}
\put(0,72){\footnotesize x}
\put(77,4){\footnotesize y}

\put(106,8){\scriptsize z}
\put(101,72){\footnotesize x}
\put(178,4){\footnotesize y}

\put(207,8){\scriptsize z}
\put(202,72){\footnotesize x}
\put(279,4){\footnotesize y}

\put(308,8){\scriptsize z}
\put(303,72){\footnotesize x}
\put(379,4){\footnotesize y}

\put(409,8){\scriptsize z}
\put(404,72){\footnotesize x}
\put(480,4){\footnotesize y}
\end{picture}
\caption{\label{fig:doublePlate} Electric field distributions of QNMs in a coupled parallel plate configuration. (a)  $\hat{E}_x$-field of the fundamental symmetric QNM ($\lambda=891+68\text{i}$~nm). (b) $\hat{E}_x$-field of a higher harmonic anti-symmetric QNM ($\lambda=622+14\text{i}$~nm). (c) -- (e) $\hat{E}_x$, $\hat{E}_y$, and $\hat{E}_z$-fields of the fundamental anti-symmetric QNM ($\lambda=1034 +34\text{i}$~nm)}
\end{figure*}
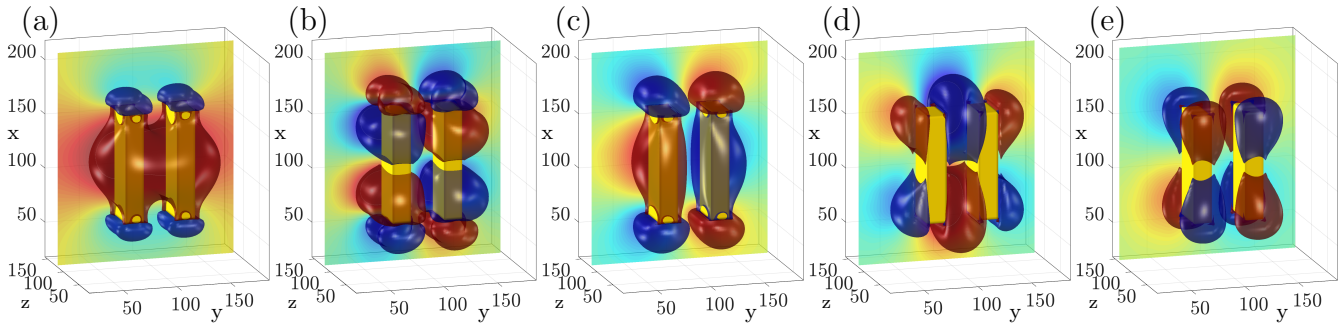

To demonstrate that the Lanczos reduction technique can also handle configurations in which multiple QNMs contribute to the SD rate, we compute the Purcell factor of a quantum emitter that is placed 10~nm above a $102~\text{nm} \times 40~\text{nm} \times 20~\text{nm}$ nanoplate as illustrated in  Fig.~\ref{fig:singlePlate}(c). The wavelength of interest now runs from $0.5~\mu$m to $1.2~\mu$m so that the contribution of higher order QNMs can be investigated. The Purcell factor is computed using the ROM of Eq.~(\ref{eq:SD_rate_ROM}) and the converged model is shown in Fig.~\ref{fig:singlePlate} (solid line). Without any {\it a priori} mode selection, a low rank expansion in QNMs can now be obtained by ranking the individual contributions of the approximate eigenpairs $(\theta^{[m]}_j,\sV_m\sz^{[m]}_{j})$ to the Purcell factor. For this configuration, we find that essentially only three QNMs are required to accurately describe the Purcell factor on the considered wavelength interval. The resulting three-term QNM expansion is shown in Fig.~\ref{fig:singlePlate} along with the contribution of each QNM separately. The real parts of the $\hat{E}_x$ fields of the contributing QNMs are shown in Figs.~\ref{fig:singlePlate}(a-b) for the higher order modes and in Fig.~\ref{fig:singlePlate}(d) for the fundamental QNM.

%
%
Finally, to show that QNMs in configurations consisting of multiple dispersive nanoresonators can be determined as well, we place two copies of the golden nanoplate next to each other such that the largest faces are parallel. The distance between the plates is 38~nm. This configuration supports anti-symmetric and symmetric resonances, where the wavelength of the anti-symmetric resonance is larger than the wavelength of the symmetric resonance in accordance with the theory of electronic oscillators. In particular, the wavelengths of the fundamental anti-symmetric and symmetric resonances are $\lambda=1034+34\text{i}~$nm and $\lambda=891+68\text{i}~$nm, respectively. Figure~\ref{fig:doublePlate}(a) shows an isosurface plot of $\text{Re}(\hat{E}_x)$ of the fundamental symmetric QNM, whereas isosurface plots of $\text{Re}(\hat{E}_{x/y/z})$ of the anti-symmetric resonance are shown Figs.~\ref{fig:doublePlate}(c) -- (e). Finally, a higher harmonic anti-symmetric resonance is depicted in Fig.~\ref{fig:doublePlate}(b).

In conclusion, we have shown that the symmetry of Maxwell's equations can be used to effectively compute QNMs of three-dimensional arbitrarily-shaped dispersive nanoresonators. A mimetic discretization of the first-order Maxwell equations for dispersive media leads to a large-scale discretized Maxwell system that is symmetric with respect to a particular bilinear form. This symmetry property allows us to reduce the large-scale Maxwell system to a system of much smaller order via a Lanczos-type reduction process and to find QNMs that are quasi-orthonormal with respect to the bilinear form.  Moreover, we have presented a new closed-form reduced-order model for the SD rate of a quantum emitter that is parametric in wavelength meaning that a single model approximates the SD rate over a complete wavelength interval of interest, i.e. the model allows for wavelength sweeps. This feature is important in many applications in quantum optics, where the SD rate is controlled and optimized by modifying the background configuration of the quantum emitter. Specifically, for each background realization a single ROM provides an SD rate response over a complete wavelength interval of interest, which can significantly speed up the design and optimization of the resonating environment. Furthermore, the ROM does not require an \emph{a priori} expansion of the electric field in terms of QNMs. It is not necessary to determine beforehand which QNMs contribute the most to the electric field at the dipole location. In fact, which modes actually contribute on a given wavelength interval can be determined \emph{a posteriori} from the reduced Lanczos system and the corresponding converged ROM by ranking and superimposing the most contributing modes until a specified error criterion is met. In this manner, the ROM for the SD rate gives us control over the error that is introduced when a subset of QNMs is used to approximate the SD rate of a quantum emitter.

%
%
\bibliography{QNMandSDRviaLanczos2}

\end{document}